# Where are your Manners?
# Sharing Best Community Practices in the Web 2.0

Angelo Di Iorio
Department of Computer Science
University of Bologna
Mura Anteo Zamboni 7
40127 Bologna, ITALY
+390512094871

diiorio@cs.unibo.it

Davide Rossi
Davide Rossi
Department of Computer Science
University of Bologna
Mura Anteo Zamboni 7
40127 Bologna, ITALY
+390512094978

rossi@cs.unibo.it

Fabio Vitali
Fabio Vitali
Department of Computer Science
University of Bologna
Mura Anteo Zamboni 7
40127 Bologna, ITALY
+390512094872

fabio@cs.unibo.it

Stefano Zacchiroli
Stefano Zacchiroli
Laboratoire PPS
Université Paris Diderot
UMR 7126
175 Rue du Chevaleret,
75013 Paris, FRANCE

zack@pps.jussieu.fr

## ABSTRACT
The Web 2.0 fosters the creation of communities by offering users a wide array of social software tools. While the success of these tools is based on their ability to support different interaction patterns among users by imposing as few limitations as possible, the communities they support are not free of rules (just think about the posting rules in a community forum or the editing rules in a thematic wiki).

In this paper we propose a framework for the sharing of best community practices in the form of a (potentially rule-based) annotation layer that can be integrated with existing Web 2.0 community tools (with specific focus on wikis).

This solution is characterized by minimal intrusiveness and plays nicely within the open spirit of the Web 2.0 by providing users with behavioral hints rather than by enforcing the strict adherence to a set of rules.

## Categories and Subject Descriptors
H.3.4 [**Information Storage and Retrieval**]: Systems and Software - *World Wide Web (WWW);* J.4 [**Social and behavioral sciences**] – Sociology.

## General Terms
Languages

## 1. INTRODUCTION
The social software tools of the Web 2.0 [22] (blogs, wikis, forums, folksonomies, ...) have fostered the creation of many web-based organizations by offering groups of users a basic set of community-building tools. Since those tools are targeting the wider possible audience, they do not assume any specific organizational structure, thus promoting "emergent coordination", i.e., the dynamic shaping of an organizational structure out of the usual practices of a community.

As time passes, these tools, originally conceived to support bazaar-like organizations (wikis, for example, have been originally designed to support software development teams), are drawing more and more interest from cathedral-like organizations (as is the case of Enterprise 2.0). And, of course, of anything in the between.

In this context we see the rise of new requirements related to the rules of the specific organizations. After all, even non-structured organizations have rules (possibly "soft" rules, such as best practices). A community forum may have posting rules, a thematic wiki may have formatting rules, and so on. Of course most of the times, these rules are checked by a human reviewer (or moderator, or editor) since they are content-related rules (e.g., "avoid political content", "do not post copyrighted material", etc.), but other times the rules are related to structure ("do not post images larger than X", "always include the abstract at the beginning of the document", etc.). In this second case it would be beneficial to use some automatic tool for checking the adherence of the published contents to the organization's rules.

The obvious solution to this problem is extending the existing tools to enforce the adherence to the organization's rules. But, in a Web 2.0 context, this solution has several drawbacks. First of all, even when the tools' sources are available (which is often the case since most social software tools are open source software) it makes little sense to create an organization-specific branch enforcing the local rules, as this would mean loosing the chance to integrate all future fixes and updates of the software. Moreover, most Web 2.0 tools are not run by the organization itself but are often made available as services from other providers (such as yahoo groups, Google docs and so on), following the software-as-a-service approach (which is integral part of the Web 2.0 philosophy). In all these cases, modifying the tools' code is not an option.

But, as *mashups* (i.e., mixing existing services to create new ones) and *layered services* (i.e., services that make use of other available services, like a meta-search engine) show, modifying the code is not the only way to personalize the behavior of social software tools.

In this paper we present a framework for supporting the application of the best practices of an organization by layering on top of existing social software tools a rule-checking service that can help the members of the community. This solution does not enforce the users to meet the rules but rather, in a way that is consistent with the "open approach" of the Web 2.0, it is based on dynamic annotations that highlight the suspicious content, allowing the user to fix it at a later time.

Since best practices can be shared among different organizations and actors engaged in different roles within the same organization can have interest in a different set of rules, a relevant feature of this framework is the ability for each single user to activate a specific set of rules among the many different ones that can be shared across different organizations.

This paper is structured as follows: Section 2 investigates scenarios where users would benefit from automatic and customized annotations of web content, focusing on wikis; Section 3 introduces our rule-based solution and compares it with similar approaches; Section 4 discusses implementations issues, while Section 5 presents some related works.

## 2. THE NEED FOR EXTERNALIZED AND PERSONALIZED CONTENT

While some community-specific best practices can be neither enforced nor checked automatically, many others can. The framework we are proposing focuses on the latter classes, and deliberately avoids to enforce them, rather enabling to verify them, pointing out where they have been violated. Our focus are wikis, the ingredient of Web 2.0 representing the more communitarian evolution of content management systems (CMSs), even though our framework can be extended to other CMSs, including blogs. With wikis however, yet another reason for not enforcing community best practices exist: doing so would sensibly diminish the editing freedom and ease of access of contributors, which is one of the key ingredient of the success of wikis. In many communities, members would prefer content contributions violating the best practices (which can be made compliant in a second moment, possibly by a different contributor), than no contribution at all. The option suggested and embraced in this paper, is to give the contributor all the appropriate feedback related to best-practice violations, so that they are easily spotted and are more likely to be promptly fixed.

Note that strict enforcement of best-practices is still possible, with few modifications, in our approach (basically we would need to change the wiki workflow preventing users to save invalid pages; the same verification would be performed by the same components, called validators, we envisioned in the current architecture). On the other hand, we strongly believe such *light* enforcement must be preserved. Any strong enforcing process, although useful in many contexts, would distort the nature itself of a wiki. In that case, if we really need *hard-constraints*, we prefer to use a "traditional" and more controlled CMS.

Several use cases of automatically checkable best practices on wiki content can be made. One of the simplest is verifying that pages representing conceptually similar entities (i.e., instances of the same class in an ontology) have similar structure, such as verifying whether all pages in Wikipedia about countries, or music records, or animal species, conform to a common given structure. Frequently, small software-oriented wikis show similar situations: pages describing software releases, bug reports, feature requests, plugins, may be better used if conforming to a common structure. Manually checking it quickly becomes tedious and benefit from automation.

Large wiki deployments such as Wikipedia have some technical solutions to these kind of problems by the mean of functional templates [4]. Such a solution is on the one hand too constraining for users (it is quite a task to compare the actual page markup with the rendered page), and on the other this only offers a mitigation of the problem, since there is no way, for example, to force all users to use the same template or to properly set all the fields of a template.

The analogy with templates brings to a generalization of the above use case, in which (at least parts) of the desired best practices can be encoded as templates (pages in the wiki itself) with editing holes to be filled when instantiating the template. What is desirable in such cases is that template instances, usually created by copying and pasting the raw template to new pages, do not diverge "too much" from the originating template. "Too much" is usually defined as only allowing instances to provide content for the editing holes, without modifying other parts of the markup. Such checks, not possible in state of the art wikis, can be externalized and implemented using our framework, where users visiting instances can be notified of excessive dissimilarities between the page they are looking at and its original purpose (embodied by its template).

A more complex use case is even more interesting in the context of this paper: the handling of source code snippets in technical wikis. While at first glance it can seem a narrow use case, there are several reason to reason about it. The first reason is historical, since the first wiki was in fact meant to share code snippets for software enginering/design purposes [23]; we want to show how the initial wiki purposes can be improved by automatically checking best-practice applications.

The second reason is an observation about the current diffusion of wikis, which is almost ubiquitous in support and talk sites about software and software projects. The third and final reason is that also large, non-technical, general-purpose wikis such as Wikipedia have some support for source code snippets, most of the time in the form of syntax highlighting. Code snippets are often tagged as such, with also a declaration of the implementation language, to benefit of ad-hoc layout support.

Currently, support for source code snippets is simply presentational: we find syntax highlight, and font typeset, and no additional checks on their quality or uniformity with other snippets of the very same wiki. Our framework would enable:

- *Coding guidelines.* Most languages have style guidelines that ease basic reading (indentation, naming convention, ...), which are not enforced by compilers, but considered good style by competent programmers. Using our framework the community can provide checkers for adherence to styling guidelines, both for language-wide guidelines and for community-specific guidelines (e.g., for the community of developers of a single big open source project)

- *Syntactic correctness.* Usage examples for libraries are a common use case of code snippets on the web and on wikis. Usually there is no guarantee that a given snippet is syntactically correct, as at best wiki engines perform syntax highlighting which is usually based on coarse-grained language grammars. With our framework, user communities can define their own syntactic checkers (possibly by plugging into real-life compilers) and have on-the-fly syntax verification of code snippets as soon as the page is read. Even though this kind of checks would mark as invalid any example of code shorthand and placeholders (such as "…") this would be a minor issue in the handling of correct checking, and special code would be implementable to skip them.

- *Testing.* Specific development communities can push snippet code checking even further. Knowing for example that the snippets are meant to be used with a given language and library, the external checker can actually compile and run snippets (of course in a sandbox, for obvious security reasons), pointing out directly on the page any snippet that is somehow broken. This would not only help the casual reader in avoiding losses of time with non-functional test cases, but also help the page maintainer to spot which code samples have been recently broken (e.g., due to software releases) and need to be fixed.

Strong analogies can be found between this specific case and any other scenario where users are interested in checking multiple and heterogeneous requirements over a wiki content. A point is crucial: the verification process is not meant to be embedded and shared by all the wiki users, but requires to be personalized and computed on-the-fly when a given page is accessed.

## 3. RU ANNOTATIONS: EXTERNALIZED PERSONALIZATION FOR WEB 2.0 CONTENT

It is not difficult to implement a wiki supporting multiple access policies to the same content. In fact, some tools already provide such feature (such as Twiki [16]) or plan to support it very soon. That kind of personalization is hard-coded in the wiki itself: users have to be registered and associated to a given profile for accessing the personalized content, wiki pages have to be properly marked-up and obviously the support for personalization has to be part of the wiki engine code.

On the other hand, our goal is to help users get more functionality on the content they generate without interfering with the original wiki workflow. The wiki remains a "passive" content management system, whose users are unaware of the external personalization process. *Externalized personalization* is then a first keyword to describe our approach.

Different types of *externalized personalization* can be envisioned: personalized content can be a completely different resource obtained by transforming the original one; or a filtered record of information items derived by removing non-relevant data; or even a version of the original content annotated with some extra data.

We are particularly interested in the last category, that of adding new data as *annotations* on the original content of the resource. Before going into details of our approach, let us clarify the meaning of the word "content" in this setting. Heterogeneous "content" in fact exists in the Web 2.0 era, delivered through heterogeneous platforms (from text to images, from animations to videos). Although we focus on "textual content of a wiki page", considering wikis as the most representative and flexible authoring systems in the Web 2.0 panorama, our analysis can be directly extended to blogs or similar systems. These systems are in fact characterized by an open editing approach which allows users to freely and easily edit textual content. Nevertheless, it is not difficult to extend our ideas to systems dealing with "structured textual content" like address books and bookmarks. In that case, it would be even easier to add annotations, since the automatic annotation process would run on well-defined data structures. An interesting difference exist between text and multimedia: annotating multimedia content would require to use specific tools and techniques dealing with multimedia metadata and encoding. Nevertheless the architecture we propose, based on decoupling the verification and annotation processes, can be generalized to manage such content too. However, annotations on multimedia content are currently out of the scope of our research.

### 3.1 Rule-based, User-defined Annotations

We are instead interested in adding new data as *annotations* on the textual content of a web page. In particular, our goal is to provide users the support for *automatically annotated content*. The basic idea is to let users to "declare" a set of properties they want to verify on some content or a set of filters they want to apply. The annotated page is the result of a conversion process, where annotations are not created manually by a user but, rather, by an automatic agent processing the user's declarations on the input page. Such declarations do not have to be embedded in the source code of a page but can be retrieved and processed on-the-fly when accessing that page. Moreover annotations can be strictly personal, or shared by a group or by the whole community.

These considerations lead us to design an idea for web content annotations we called *RU Annotations*, expanded in "Rule-based, User-defined Annotations".

In the scenario we envision, each user defines rules to verify and filter wiki pages and dynamically associates these rules to the accessed pages. Whenever a page is displayed, the associated rules are processed over its content by invoking a specific agent. The result of this process is merged with the original content and displayed in the browser. Note that the original content remains unmodified in the origin server, while users access a personalized and automatically annotated view of that page.

It is evident that technical issues have to be addressed to implement such a solution. The most important will be investigated in Section 4. Here we want to highlight some additional advantages of *RU Annotations* beyond the use cases discussed before:

- Content filters nowadays work on full pages (or even full sites), and pretty legitimate documents, posts or comments are not accessed because of the unfortunate use of a single word within it. RU Annotations could prevent the single offending string to be displayed, while the full message would still be readable.
- Disabled users trying to read through non-accessible pages often discover their impossibility to read through their content halfway through the page. RU Annotations could provide immediate feedback on the accessibility level of the individual resource, leaving the user decide whether to continue reading or not.
- Automatic reputation evaluators could be constructed on top of web resources without changing a single byte in their content. As such, unreliable, badly-reputed or fringe content could be therefore correctly annotated by users not interested in fringe communities or unverified assertions.

### 3.2 From annotated view to RU annotated view

*RU Annotations* are strongly connected with the analysis of the wiki editing process and in particular of the wiki annotations presented in [1]. Authors first introduced the concept of wiki ANNOTATED VIEW to indicate the possibility of displaying a wiki page enriched with some extra-information which are not directly written by the author but dynamically added by the wiki engine when rendering that page. These annotations would not prevent the wiki users to create the resources, but would constitute a continuous feedback with regard to the perceived or computable "correctness" of the content according to locally specified rules. These extra information were called "Light Constraints" (or *LC Annotations*), since they would not obstacle any wiki-specific activity, but just add to the smoother functioning of the system.

The "externalization" and "user-based customization" of such operation leads us to shape the concept of "Rule-based User-defined Annotations". Let us briefly compare RU Annotations with LC Annotations.

The operation of ANNOTATED VIEW (and the CONDITIONAL SAVE, i.e., saving content only after verifying some of its properties) came with the idea of Light-Constraint Wikis. A Light-Constraint wiki is an enhanced wiki managing constraints over the wiki content and allowing users to (i) define rules and constraints which should be verified on that content, (ii) view annotated pages with detailed report on such verification process, and (iii) verify content before saving it.

Explicit and implicit constraints exist in the wiki context too, although wikis seem to be completely free and open. For instance, wiki authors frequently need to create sets of pages with the same structure (like pages for courses/professors in a university wiki), or need to write correct in-line fragments in a non-wiki language (like MathML or LaTeX formulas, which must be correct), or need to keep consistent data structures in multiple wiki pages (like in the lists used in connected Wikipedia pages) and so on. On the other hand, wikis have to keep their completely open and free editing model. Thus, an interesting question arises: is it possible to integrate some form of constraints in the free wiki editing process?

The "*light-constraints*" approach relies on encoding these requirements as soft constraints that can be (temporarily or not) violated, without inhibiting proper wiki runtime behavior. The idea is to let authors to declare constraints on each page and to provide users a detailed report on the constraints' verification process, when displaying or saving that page. The key point is the lightness of such constraints, which can verified but also ignored by users: light-constraints do not distort or weaken the wiki workflow. A general architecture, adaptable to different wiki implementations, can be instantiated to manage light-constraints. It relies on a strong distinction between the actual wiki engine and a set of modules, in charge of verifying the respect of light constraints associated to the pages. These modules are called *validators* and envisioned as internal processes or external services or pluggable sub-components. The introduction of Light-constraints and validators transforms the basic VIEW operation into an ANNOTATED VIEW.

The validation report is clearly visible and separated from the original content. This is a first important feature that makes LC and RU annotations very similar: both these approaches add an extra-layer to the original content. This layer is not directly created while authoring the page, rather it is an optional and separated resource added on-the-fly. It does not mean that annotations are necessarily appended at the end of the page. On the contrary, reports can be "localized", i.e., composed of textual messages bounded to particular characters in the wiki markup in order to be easier to spot and fix.

The second important point is that both these approaches only manage "light" constraints: in fact annotations can also be deactivated and ignored by the users. The presence of the aforementioned extra-layer does not interfere with the wiki workflow and, more importantly, does coexist with the free and open wiki editing model. Wiki authors keep on being able to fully modify the wiki content, without taking care of validation and rule-based filtering. These operations will be performed *a posteriori* in a complete transparent manner.

On the other hand, important differences can be outlined. First, LC and RU annotations differ for the intrusiveness and interaction with the wiki workflow. In the LC architecture, rules and constraints are defined *a priori* during the editing process and each page directly contains references to the associated validators. In the RU scenario, on the contrary, rules and constraints are applied *a posteriori* when a user accesses a given page.

Second, LC annotations are displayed whenever a given page is accessed and all users view the same annotations. On the other hand, RU annotations are personalized and displayed only when a given user accesses that page. Different views and annotations can be applied to the same page, and different users can be unaware of the constraints and rules of others.

These differences derive from the nature and scope of LC and RU annotations. The first ones are meant to encode constraints shared by the whole community and implicitly developed during the authoring process. They are particularly useful for verifying the intrinsic quality of the wiki content and lightly enforce the fulfillment of community-based and content-based requirements. On the other side, RU annotations are particularly useful when the validation process does not involve all the wiki users, or needs to encode some specific sub-community best-practices.

Moreover RU annotations do not require users to embed validation information within the code of a page and allows users to also annotate wiki site they are not entitled to modify. As happened for early hypermedia systems [5] or web annotation projects [6] even read-only resources can be annotated since the annotations are stored externally or computed on-the-fly.

The possibility of sharing annotations is another point worth being discussed. It is easy to envision policies for reducing the visibility of some LC annotations to subgroups: such approach requires important modifications to the code of the wiki platform. On the other hand, RU annotations can be shared by simply sharing the rules generating them, and group-based policies can be supported without impacting the wiki code.

Finally, LC or RU differ from the implementation perspective. LC annotations require implementers to modify the internals of the wiki while RU annotations are not intrusive at all. On the other hand, the first approach is a bit easier to be implemented having full control on the internal modules of the wiki.

## 4. FROM DESIGN TO IMPLEMENTATION
Several alternative possibilities exist to implement the proposed framework. In this section we analyze these alternatives, outlining the technical solutions to support them, and present a proof-of-concept implementation based on a proxy architecture.

## 4.1 Global architecture, validators, and rules: basic design issues
One of the first choices that have to be made is related to the overall architecture of the system. Regardless of the actual implementation of each component, our architecture can be generalized as in fig. 1.

The user interacts with a common web browser and normally requests a page, by typing a URL or following a link. That page is retrieved (unmodified from its origin server) and annotated by an intermediate component we call "Annotator". The Annotator includes a module firing rules (associated with the URL) and a validation agent checking the content against those rules. Once annotations are merged into the original page, the final result is returned to the user.

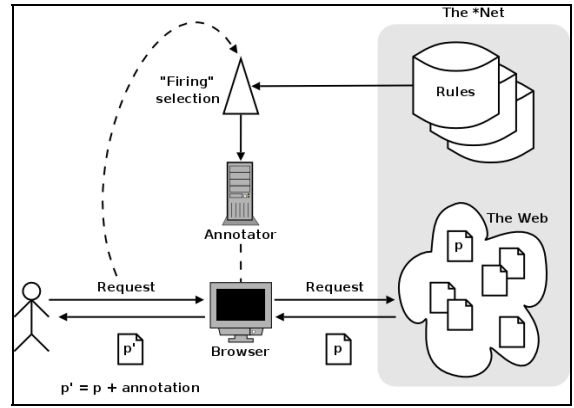

**Fig 1: The overall architecture for RU Annotations**

Many architectural issues arise from this scenario. First of all, we need to clarify the position of the Annotator (the dashed line in the picture indicates the fact that such component is not strictly required to be in the browser). Two solutions are possible: a man-in-the-middle approach, and a client-side approach. The man-in-the-middle approach can be implemented by using a filtering proxy (possibly a personal proxy, hosted in the client's machine) or with a special web application acting as a reverse proxy. Obviously the first solution is more transparent (users behave like they are connecting directly with the origin server) but is more intrusive (users have to change the preferences of their browsers in order to use the proxy). The client-processing approach can also be implemented with two main techniques: bookmarklets and browser extensions. Bookmarklets are the least intrusive solution. They follow a distinct opt-in approach, in which users have to explicitly activate the bookmarklet in order to add the annotations to the presented document. Browser extensions are more flexible: they can run either transparently or on-demand and have fewer limitations with respect to bookmarklets. Obviously they require a supported browser and have to be installed (but this can be a matter of just few clicks).

Another relevant question is related to the validation agent: where is it hosted? Also in this case we have different possible solutions: is the agent remote or is it running in the client's machine? Furthermore: a local agent, in a client-side processing scenario, can be integrated inside the bookmarklet/extension or can run as a separate software component. Consider that, in any case, the agent has to access the shared rule repository, which is an essential requirement in our approach. It might be a public repository or an intranet resource. Also consider that an implementation could support several validation agents.

A last issue is related to the validation agent: how does it operate? Users subscribe some of the rules from the shared repository. A rule is characterized by a firing part (used to decide if the rule applies to the current document) and an active part (in which the document is analyzed and possibly the annotations are produced). The rule-checking agent has to fire all the relevant subscribed rules and return the annotations to the caller, which is responsible for alayering them on the page.

Several techniques can be used to implement both the firing part and the active part. A simple approach is to use, for the firing part, regular expressions that match the URL of the document and/or XPath expressions that match its contents. XSLT is the more straightforward solution for the active part. As stated above it is well possible that several agents, based on different technologies, exist in the same implementation. Notice that the ability to express complex, semantic rules with this approach is strongly dependent on the semantic annotations included in the document. Our solution, in fact, operates on the displayed document, not on an intermediate format (the one managed by the wiki engine, that is later transformed in HTML). It is then essential that the semantic information available from within the system emerges also in the HTML page; this is now possible with the adoption of techniques like microformats [2] or RDFa [3].

An orthogonal aspect is the language used to express the active part, which has to check the document and annotate it with the outcome. While several interesting trade-off, e.g., about usability, have been previously discussed regarding the language choice [1], we observe that such a choice is independent from the overall framework architecture. We plan to experiment with a wide range of languages (from the implementation language of the wiki engines themselves, to ontological languages where semantic wikis can be assumed), the actual choice is outside the scope of the present work.

### 4.2 A proof-of concept-implementation

It is clear that several options exist for the implementation of the proposed framework. All have pros and cons, they have to be evaluated with respect to the specific context in which the solution has to be deployed (Is it possible to assume that all the members of the community use a specific browser? Can we ask them to install a software package in their computer? Do we prefer an opt-in approach?). We experimentally implemented a proxy-based solution (modifying an existing filtering proxy we developed for another project) in which the checking agent is integrated with the proxy and accesses a simple, read-only, remote shared rules repository (with rules based on the regular expressions/XPath and XSLT approach hinted above). The rules are used to check a wiki page with respect to its original template (similarly to the example given in section 2). This proof-of-concept implementation showed that our framework can be easily instantiated. The full potential of the system, however, can be reached only by defining mechanisms to share the rules in the repository (or repositories).

### 5. RELATED WORKS

The external annotations of web content are not new in the literature. A lot of systems have been presented in the hypertexts and hypermedia community since the early days of the WWW. The very first browser NCSA Mosaic [15] allowed users to created personal and locally stored annotations to web pages. Later, researchers focused on the possibility of adding external annotations too: personal notes were stored on external linkbases and added on the fly to any web page. The Arakne framework [5], for instance, provided users a powerful interface to add annotations and links to any web page.

Standard languages have also been proposed in this field. The Annotea project [6] is a W3C effort to standardize the process of creation, retrieving and dissemination of external annotations based on RDF but easily extensible. Although powerful Annotea implementations exist, the protocol did not succeed as expected. Similary, XLink is a very powerful W3C standard for advanced XML linking which also allows users to express external annotations and comments. As example, XLinkProxy [7] is a proxy-based application which support users in authoring external XLink annotations and dynamically adds these annotations to web pages.

More recently, annotations are gaining great importance in the Web 2.0 scenario. For instance, Diigo [8] is a powerful knowledge platform allowing users to share their comment on any web page. Users are required to join the Diigo community, a toolbar can easily be installed to their browser and supports them during the annotations authoring, searching and collecting process. A similar approach is implemented by Sharecopy [12], whose website allows users to download a multi-browser toolbar for external annotations. Thousands of other web2.0 social annotation tools could have been listed here, all based on Javascript and DOM manipulations. Most of them are specialized for specific purposes like WizLite [17] tailored for text-fragment highlighting, or Trailfire [18] for annotating and sharing customized path among different web pages or GoogleNotebook [19] to collect data while surfing the web using a powerful tool in the Google framework.

All these systems support users in the *manual* annotation of web content and allowing surfers to highlight fragments and add localized and sticky notes. On the other hand, RU annotations are generated by a verification process: users only declare properties and constraints to be satisfied and an automatic agent produces the final output.

The automatic extraction of information from web pages has also been widely researched. In [13] authors proposed to exploit visual clues and similarities among different pages in order to re-build the logical structure of a web page. Similar layout-based approaches are opposed to the DOM-based analysis techniques, which exploits the automatic recognition of patterns in the HTML organization of a web page. In [14] authors presented a tool which highlights the role of each fragment in a web page through a pre-filtering phase in charge of selecting images, objects, links, logos and a post-filter phases in charge of interpreting chunks of text and structured data.

Tools for automatic analysis of web content have also been developed by the researchers interested in crawlers and RSS harvesters and aggregators. They are mainly based on heuristics and statistical analysis but cannot be fully described here due to the space limits of this paper. All these solutions implement very powerful and general techniques working on different content. On the other hand, RU annotations are targeted to a specific context so that more precise results can be achieved in the automatic analysis phase.

All these efforts share a common background with RU annotations: the need of integrating multiple and heterogeneous sources of information through a post-processing approach. In [20] authors propose to automatically create a federation of Web 2.0 websites by exploiting multiple tagging and on-the-fly analysis of content. Similarly [21] presented a tool to automatically extract annotations from web pages and outline semantic relations between content-related pages. Although their approaches are mainly focused on metadata, these efforts showed how very heterogeneous web resources can be checked and unified by a transparent and *a-posteriori* process as proposed by RU annotations.

## 6. CONCLUSIONS

In this paper we have presented a framework to add rule-based annotations to resources that would allow web sites, organizations, users or third party to add automatic annotations to wiki resources without interfering with the normal workflow of the origin application but that can be used for on-the-fly decisions about the content or the validity or the appropriateness of the page before displaying it.

We envision scenarios in which communities share knowledge about their best practices in the form of a set of shared rules that each user can independently decide to subscribe (because of their personal interests or because of the role they play in the organization since it is well possible that users with different roles are interested in checking different best practices). By setting an option in their browser (when using extensions or a filtering proxy) or by clicking a button (to activate a bookmarklet) the users can ask the system to check the subscribed rules and annotations are layered on top the current document when violations occur.